\documentclass[aps,pra,twocolumn,showpacs,superscriptaddress]{revtex4}
\usepackage{graphicx}
\usepackage{amssymb}
\usepackage{amsmath}
\usepackage{bm}

\begin{document}

\title{Role of thermal friction in relaxation of turbulent Bose-Einstein condensates}

\author{Joon Hyun Kim}
\affiliation{Department of Physics and Astronomy, and Institute of Applied Physics, Seoul National University, Seoul 08826, Korea}
\affiliation{Center for Correlated Electron Systems, Institute for Basic Science, Seoul 08826, Korea}

\author{Woo Jin Kwon}
\affiliation{Center for Correlated Electron Systems, Institute for Basic Science, Seoul 08826, Korea}

\author{Y. Shin}\email{yishin@snu.ac.kr}
\affiliation{Department of Physics and Astronomy, and Institute of Applied Physics, Seoul National University, Seoul 08826, Korea}
\affiliation{Center for Correlated Electron Systems, Institute for Basic Science, Seoul 08826, Korea}

\begin{abstract}
In recent experiments, the relaxation dynamics of highly oblate, turbulent Bose-Einstein condensates (BECs) was investigated by measuring the vortex decay rates in various sample conditions [Phys.~Rev.~A $\bf 90$, 063627 (2014)] and, separately, the thermal friction coefficient $\alpha$ for vortex motion was measured from the long-time evolution of a corotating vortex pair in a BEC [Phys.~Rev.~A $\bf 92$, 051601(R) (2015)]. We present a comparative analysis of the experimental results, and find that the vortex decay rate $\Gamma$ is almost linearly proportional to $\alpha$. We perform numerical simulations of the time evolution of a turbulent BEC using a point-vortex model equipped with longitudinal friction and vortex-antivortex pair annihilation, and observe that the linear dependence of $\Gamma$ on $\alpha$ is quantitatively accounted for in the dissipative point-vortex model. The numerical simulations reveal that thermal friction in the experiment was too strong to allow for the emergence of a vortex-clustered state out of decaying turbulence.
\end{abstract}


\maketitle

\section{Introduction}
In a superfluid where vorticity is quantized, a turbulent flow is formed with a complex tangle of many vortex lines, which is referred to as quantum turbulence (QT)~\cite{Skrbek12,Tsubota13}. QT has been studied for many decades in superfluid helium, leading to an intriguing comparative study between QT and classical fluid turbulence~\cite{Vinen020610}. Atomic Bose-Einstein condensates (BECs) are actively considered a new system for QT because of recent experimental advances in generating and imaging quantized vortices~\cite{Henn09,Neely10,Freilich10,Kwon14,Neely13,Moon15,Wilson15,Seo16}. QT is also discussed in the context of far-from-equilibrium quantum dynamics, which is one of the frontiers of current quantum gas research~\cite{Eisert15}.

Many of the recent works on QT in BECs address the decay of two-dimensional (2D) turbulence. The key question is whether a large-scale vortex structure emerges in decaying 2D QT. This phenomenon is known as the inverse energy cascade and is well established in 2D turbulence in a classical hydrodynamic fluid~\cite{Kraichnan75,Kraichnan80,Tabeling02}. Based on the Gross-Pitaevskii (GP) equation for the condensate wave function, many numerical efforts  were made to answer the question, but there is still no consensus on the emergence of inverse energy cascades, especially in compressible 2D QT~\cite{Numasato10,Bradley12,Chesler13,Reeves13,Billam14,Chesler14,Simula14,Billam15}. In experiments, the 2D regime was addressed by employing BECs with oblate geometry, where the vortex line is energetically aligned along the tight confining direction and thus, vortex dynamics is effectively 2D~\cite{Jackson09,Rooney11}. Neely $\it{et~al.}$~\cite{Neely13} reported tantalizing experimental evidence of the inverse energy cascade by observing vortex pinning in an annular BEC under small-scale stirring. However, Kwon $\it{et~al.}$~\cite{Kwon14} observed no signature of large-scale vortex formation in their investigation of the relaxation dynamics of highly oblate, turbulent BECs over a wide range of sample conditions.

Our interest in this paper is in the effect of thermal damping in the evolution of a turbulent BEC, which arises because of the interaction between the condensate and the coexisting thermal atoms at finite temperatures~\cite{Fedichev99,Kobayashi06,Berloff07,Madarassy08,Jackson09,Rooney10,Thompson12,Gautam14}. So far, most of the theoretical works on 2D QT in BECs have concentrated on the low-temperature regime by studying the GP equation, if any, with a small phenomenological damping constant. Some of those works predicted clustering of same-sign vortices in decaying turbulent condensates~\cite{Reeves13,Billam14,Simula14,Billam15}. Because thermal damping has a tendency to drive a vortex state to a stationary state~\cite{Billam15}, there must be an upper bound of thermal damping for observing vortex clustering. It is practically important to figure out whether the temperature requirement is achievable in current experiments.

\begin{figure*}
\includegraphics[width=14.0cm]{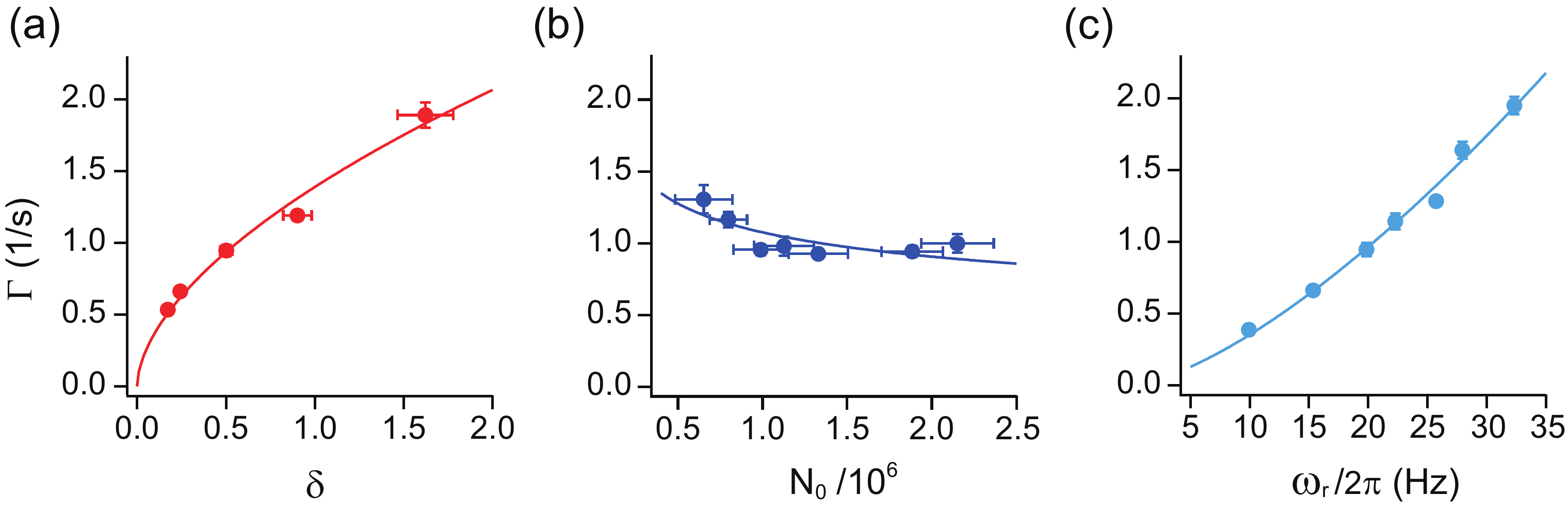}
\caption{(Color online) Vortex decay rate $\Gamma$ determined from the experimental data of Ref.~\cite{Kwon14}. $\Gamma$ is determined as the inverse of the time for which the vortex number decreases from 32 to 16.  The sample condition is specified with three parameters: (a) the population ratio of the thermal component to the condensate, $\delta$, (b) the condensate atom number, $N_0$, and (c) the radial trapping frequency, $\omega_r$, of the harmonic potential. For each measurement, the other sample parameters were maintained constant: in (a), $\omega_r/2\pi\approx 15~$Hz and $N_0\approx 1.7\times10^{6}$, (b), $\omega_r/2\pi\approx 15~$Hz and $\delta\approx 0.6$, and in (c), $\delta\approx 0.25$ and $N_0\approx 1.9\times10^{6}$. The solid lines denote power-law curve fits to $\Gamma$, yielding the exponents (a) 0.57(5), (b) -0.25(9), and (c) 1.45(8).}
\end{figure*}

In a recent experiment~\cite{Moon15}, Moon~$\it{et~al.}$ demonstrated that vortex motion in a BEC at finite temperatures is well described by mutual friction between the condensate and the thermal component~\cite{Hall56,Barenghi83,Schwarz8588,Berloff14}, and measured the dimensionless friction coefficient $\alpha$ as a function of temperature. In light of the $\alpha$ measurement, in this study, we revisit the experimental results of Kwon~$\it{et~al.}$~\cite{Kwon14} and examine the dependence of the vortex decay rate in turbulent BECs on the friction coefficient $\alpha$. We find that the vortex decay rate is almost linearly proportional to $\alpha$, and observe that the finding is supported by numerical simulations using a point-vortex model including longitudinal thermal friction. Furthermore, the vortex decay rates obtained from the simulations are quantitatively consistent with those measured in the experiment, indicating that thermal friction is the dominant dissipation mechanism in decaying turbulence. One notable observation in the numerical simulations is that vortex clustering can occur in the absence of thermal friction, i.e., $\alpha=0$, but it is easily suppressed by small friction that is much weaker than the weakest one observed in Ref.~\cite{Moon15}. This implies that there is a quite stringent temperature requirement for observing vortex clustering in decaying turbulent BECs, providing valuable guidance for experimental efforts in studying 2D QT in BECs.

In Sec.~II, we present a comparative analysis of previous experimental results for the vortex decay rate~\cite{Kwon14} and thermal friction coefficient~\cite{Moon15} . In Sec.~III, we describe our simulation study using a dissipative point-vortex model and discuss the effect of thermal friction on vortex clustering in decaying 2D turbulence in BECs. Finally, in Sec.~IV, we provide a summary of this work.

\section{Previous experimental results}

\subsection{Revisit of vortex decay rate}

In Ref.~\cite{Kwon14}, Kwon~$\it{et~al.}$ experimentally investigated the relaxation of superfluid turbulence in highly oblate BECs and observed a nonexponential decay behavior of the vortex number $N_v$, revealing many-vortex effects in the relaxation dynamics. The decay curve of $N_v$ was found to be phenomenologically well described by the rate equation 
\begin{equation}
\frac{dN_v}{dt}=-\Gamma_{1}N_v-\Gamma_{2}N_v^2,
\end{equation}
where the decay constants $\Gamma_1$ and $\Gamma_2$ were observed to have different temperature dependence. From a simple kinetic consideration, Kwon~$\it{et~al.}$ proposed that the linear and nonlinear decay terms in the rate equation are mainly attributed to the drift-out of vortices in the trapped BEC and the vortex-antivortex annihilation, respectively. Although the rate equation is useful in quantitatively characterizing the nonexponential vortex decay curve, it is not $\it{a~priori}$ clear whether the form of the rate equation is valid to represent the relaxation dynamics of turbulent BECs. Several numerical efforts were made after the experiment~\cite{Stagg15,Du15,Cidrim16,Groszek16}, but without reaching an agreement on identifying the universal decay behavior of $N_v$.

Here, we introduce a new practical measure for quantifying the relaxation speed of a turbulent condensate. We consider a situation where a condensate has 32 vortices with zero net vorticity, and define a vortex decay rate $\Gamma$ as the inverse of the time $t_h$ for which the vortex number decreases by one half. The details of the nonexponential decay behavior of $N_v$ is completely ignored in the determination of $\Gamma$, but the value of $\Gamma$ faithfully reflects the relaxation speed of the turbulent BEC. In experiment, we can prepare a turbulent condensate with $N_v>32$ and determine $t_h$ from a curve fit of the rate equation in Eq.~(1) to the measured $N_v(t)$. The reason why we set $N_v=32$ for the initial condition is that the smallest initial vortex number in the measurements of Ref.~\cite{Kwon14} was about 30 with high temperature samples.

Figure 1 shows the vortex decay rate $\Gamma$ determined from the experimental data of Ref.~\cite{Kwon14}. The sample condition is specified with $\delta=N_{th}/N_0$, $N_0$, and $\omega_{r(z)}$, where $N_{th}$ and $N_0$ are the atom numbers of the thermal cloud and the condensate, respectively, $\omega_{r(z)}$ is the radial (axial) trapping frequency of the trapping potential, and $\omega_z/2\pi=390~$Hz. In a mean-field description, the chemical potential $\mu$, condensate radial extent $R$, and temperature $T$ are given by 
\begin{eqnarray}
&\mu&=\frac{\hbar \bar{\omega}}{2} \Big(\frac{15 N_0 a}{\bar{a}}\Big)^{2/5}\sim (N_0 \omega_z \omega_r^2)^{2/5}  \\
&R&=\frac{1}{\omega_r}\sqrt{\frac{2\mu}{m}}\sim (N_0 \omega_z \omega_r^{-3})^{1/5}  \\
&k_\mathrm{B} T&= 0.94~\hbar \bar{\omega} (\delta N_0)^{1/3}\sim (\delta N_0 \omega_z \omega_r^2)^{1/3},
\end{eqnarray}
where $a$ is the $s$-wave scattering length, $\bar{a}=\sqrt{\hbar/m\bar{\omega}}$, $\bar{\omega}=(\omega_r^2 \omega_z)^{1/3}$, and $m$ is the atomic mass. From the expectation that 2D vortex dynamics would show a scaling behavior with the characteristic energy and length scales of the system, we model the vortex decay rate by $\Gamma=\gamma\delta^a N_0^b \tilde{\omega}_r^c$ where $\gamma$ is the proportionality coefficient and $\tilde{\omega}_r=\omega_r/(2\pi \times 1~$Hz). The model fit to the data of $\Gamma$ in Fig.~1 gives the exponents $a=0.57(5)$, $b= -0.25(9)$, $c= 1.45(8)$, and $\gamma= 0.92(2)$~s$^{-1}$ .

\subsection{Correlation between vortex decay rate and thermal friction coefficient}

In Ref.~\cite{Moon15}, Moon~$\it{et~al.}$ generated a doubly charged vortex in the center region of a trapped BEC using a topological imprinting method~\cite{Leanhardt02,Shin04} and investigated the long-time dynamics of the vortex state. The doubly charged vortex was split into a pair of corotating vortices and the pair separation monotonically increased over time. The pair separation evolution was consistent with a point-vortex model including longitudinal friction (see Sec.~III A), and the dimensionless friction coefficient $\alpha$ was determined from the increasing rate of the pair separation. 

\begin{figure}[b]
	\includegraphics[width=6cm]{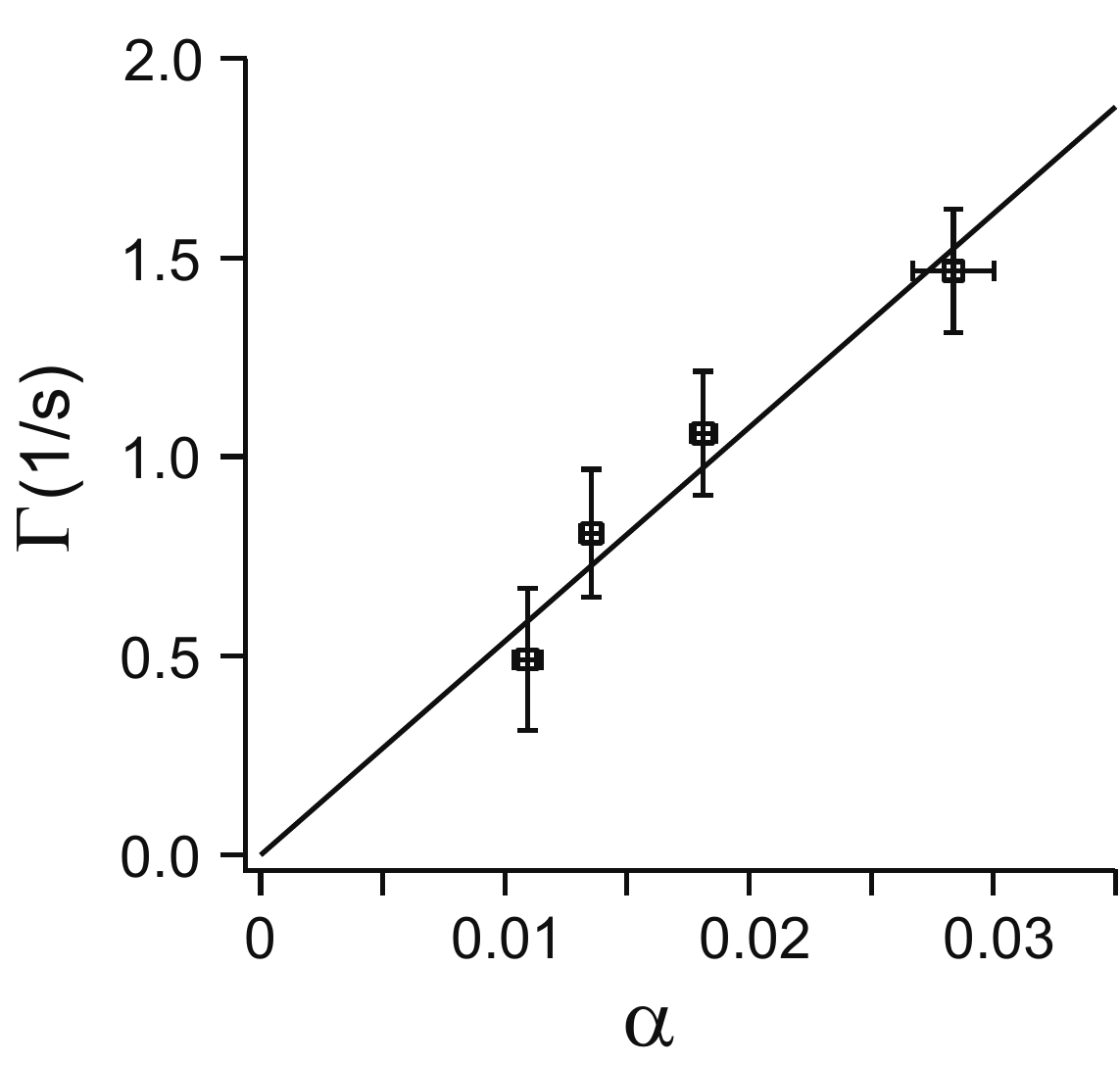}
	\caption{Vortex decay rate $\Gamma$ is estimated under the sample condition in the experiment of Ref.~\cite{Moon15} and displayed as a function of the friction coefficient $\alpha$. The solid line is a linear fit to $\Gamma$, assuming $\Gamma=0$ at $\alpha=0$.}
\end{figure}

The correlation between the vortex decay rate $\Gamma$ and the thermal friction coefficient $\alpha$ can be examined by estimating $\Gamma$ under the sample condition in the $\alpha$ measurement experiment, where $\delta$ ranges from 0.17 to 1.22, $N_0\approx 3.4 \times 10^6$, and $\omega_{r(z)}/2\pi \approx 19.7~(690)$~Hz. The sample condition is not far from the parameter window surveyed in the experiment of Ref.~\cite{Kwon14}, thus allowing to estimate $\Gamma$ from the power-law formula obtained in the previous subsection. However, the axial trapping frequency is different from that used in Ref.~\cite{Kwon14}, $\omega_z/2\pi=390~$Hz, and hence the power-law formula of $\Gamma$ cannot be directly applied to the sample condition of Ref.~\cite{Moon15}. By noting that the characteristic system parameters $\mu$, $R$, and $T$ for 2D vortex dynamics are expressed as functions of $N_0 \omega_z$ in Eqs.~(2)$-$(4), we propose a generalization of the power-law formula as $\Gamma=\gamma \delta^a N_{0,e}^b \tilde{\omega}_r^c$ by introducing the effective condensate atom number $N_0^e\equiv N_0 \omega_z/(2\pi\times 390~$Hz). Because the $N_0$ dependence of $\Gamma$ is weak, there is little room for an unexpected distortion, if any, by this generalization.

\begin{figure*}
	\includegraphics[width=15cm]{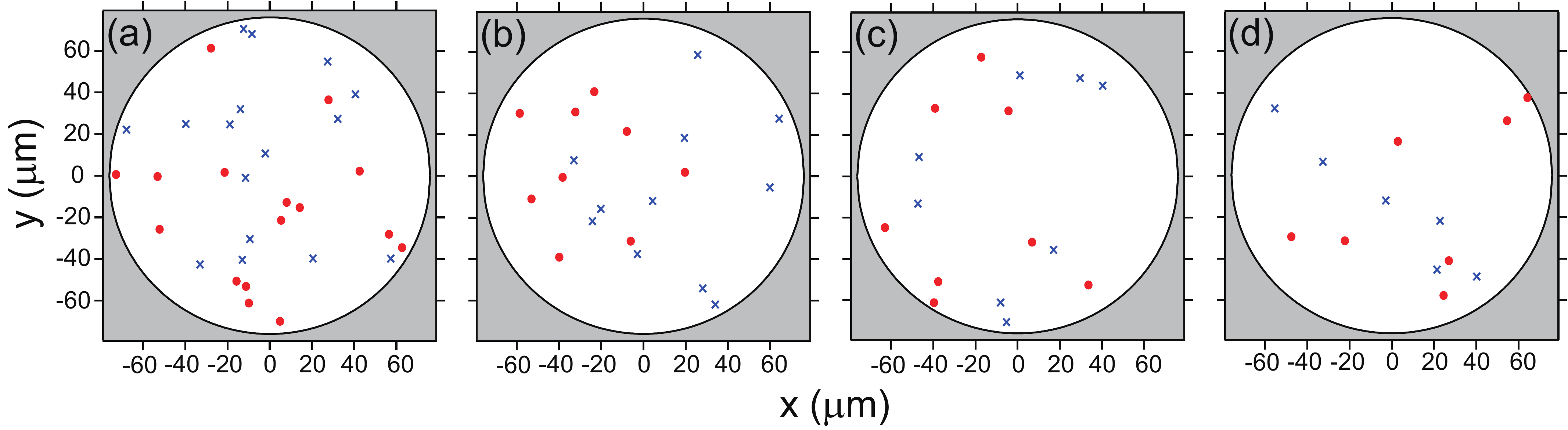}
	\caption{(Color online) Simulation of a turbulent BEC in a cylindrical flat trap using the dissipative point-vortex model.  Blue circles and red crosses denote the positions of singly charged vortices and antivortices, respectively, at $t=$ (a) $0~$s, (b) $5~$s, (c) $10~$s, and (d) $15~$s for $\alpha=0.0025$.   At $t=0$, the BEC contains 16 vortices and 16 antivortices in a random configuration.}
\end{figure*}

Figure 2 displays the estimated values of $\Gamma$ under the sample condition in the $\alpha$ measurement experiment. Because $\alpha$ represents the relative magnitude of thermal damping in vortex dynamics, it is natural to anticipate that $\Gamma$ increases with increasing $\alpha$. Interestingly, our analysis result is suggestive of a linear relation between $\Gamma$ and $\alpha$. This is quite intriguing in that the two quantities $\Gamma$ and $\alpha$ are determined separately from different vortex dynamics of trapped BECs. Because relaxation of turbulent BECs would proceed even at $T=0$ via nonthermal dissipation mechanisms such as phonon radiation~\cite{Parker04}, the linear dependence of $\Gamma$ on $\alpha$ should not be hold down to $\alpha=0$, but we believe that the nonthermal contribution to $\Gamma$ is negligible in the temperature range of the experiment~\cite{Stagg15}.

\section{Simulation}

To obtain more insights into the relation between $\Gamma$ and $\alpha$, we perform a numerical study of the time evolution of turbulent BECs using a dissipative point-vortex model. This model was successfully employed in the analysis of the long-time dynamics of a corotating vortex pair in Ref.~\cite{Moon15}, providing an essential basis for determining $\alpha$ in the experiment.

\subsection{Dissipative point-vortex model}

In the point-vortex model, a vortex is regarded as a point object that generates a circular velocity field around itself in a 2D fluid system, and its motion is determined by the velocity fields from all of the other vortices. We consider a homogeneous condensate with $N_v$ vortices in a cylindrical flat trap of radius $R$. The velocity of the $i$th point vortex is given by~\cite{Onsager49,Campbell91,Yatsuyanagi05}  
\begin{equation}
\mathbf{v}_i^0= \sum_{j\neq i}^{N_v} \frac{\hbar}{m}s_j\mathbf{\hat{z}}\times \frac{(\mathbf{r}_i-\mathbf{r}_j)}{\left|\mathbf{r}_i-\mathbf{r}_j\right|^{2}}-\sum_{j}^{N_v}\frac{\hbar}{m}s_j\mathbf{\hat{z}}\times \frac{(\mathbf{r}_i-\bar{\mathbf{r}}_j)}{\left|\mathbf{r}_i-\bar{\mathbf{r}}_j\right|^{2}} 
\end{equation}
where $\mathbf{r}_j$ is the position vector of the $j$th point vortex from the trap center and $s_j$ is its circulation in units of $\hbar/m$. The second term corresponds to the velocity field from the image vortex located at $\bar{\mathbf{r}}_j\equiv (R/\left|\mathbf{r}_j\right|)^{2}\mathbf{r}_j$ with an opposite circulation of $-s_j$, which is imposed to satisfy the boundary condition that the flow component normal to the cylindrical wall is zero.

At finite temperatures, vortex motion is affected by mutual friction arising from the relative motion of the condensate to the thermal component~\cite{Hall56,Barenghi83,Schwarz8588,Berloff14}. Assuming a stationary thermal cloud, the longitudinal friction that is proportional to $-\mathbf{v}_i^0$ gives rise to an additional vortex motion orthogonal to $\mathbf{v}_i^0$, and the resultant velocity of the vortex is given by
\begin{equation}
\frac{d\mathbf{r}_i}{dt}=\mathbf{v}_i^0-\alpha s_i\mathbf{\hat{z}}\times\mathbf{v}_i^0
\end{equation}
with the dimensionless friction coefficient $\alpha$. The time evolution of the vortex state is obtained by numerically calculating $\mathbf{r}_i(t)$ using Eq.~(6).

In the numerical simulations, we implement vortex-antivortex pair annihilation by removing two vortices of opposite circulations when they come close to each other within a certain threshold range~\cite{Simula14,Billam15}. The annihilation conditions for a vortex dipole were theoretically investigated~\cite{Rorai13}, and in our simulations we chose a critical pair separation $d_c=2\xi$~\cite{footnote1}, where $\xi=\hbar/\sqrt{2 m \mu}$ is the condensate healing length, characterizing the density-depleted vortex core size. We also allow for vortex annihilation at the wall when a vortex collides with its image vortex, which might be regarded as the drifting-out process in the trapped BEC.

Here, we need to mention the limitations of our model in describing the experimental situation. First, in the experiments, the condensate was trapped in a harmonic potential and had an inhomogeneous density distribution. The density gradient induced additional precession motions of the vortices~\cite{Middelkamp11,Navarro13}. Moreover, the local density ratio of the thermal component to the condensate varied over the sample, resulting in position-dependent thermal friction. Second, the point-vortex model solely focuses on the motional dynamics of vortices in an ideal incompressible fluid, by completely ignoring vortex-phonon interactions~\cite{Simula14,Billam15,Parker04}. In particular, vortex dynamics near the condensate boundary could be sufficiently complicated by involving density waves and surface mode excitations.

\begin{figure}
	\includegraphics[width=7.5cm]{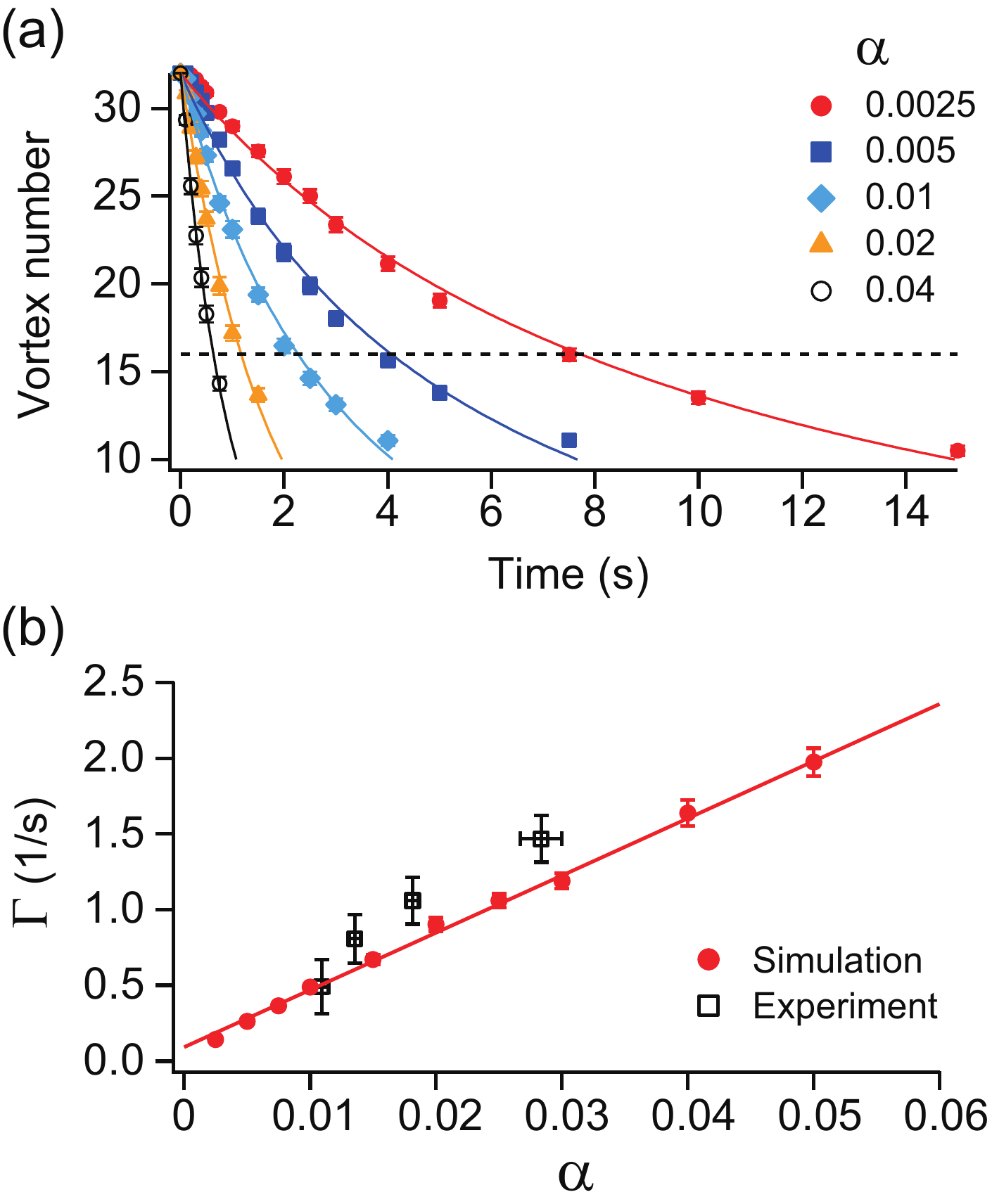}
	\caption{(Color online) (a) Simulation results of the vortex number decay curves for various friction coefficients, $\alpha$. Each data point was obtained by averaging the vortex numbers for 40 different initial states. The solid lines are nonexponential curve fits to the data using the rate equation in Eq.~(1), and the dotted line denotes $N_v=16$. (b) Vortex decay rate $\Gamma$ as a function of $\alpha$. The experimental results in Fig.~2 (open squares) are displayed together for comparison. The red solid line is a linear fit to the simulation results.}
\end{figure}

\subsection{Results}

Following the sample condition in the $\alpha$ measurement experiment, we set $R= 76~\mu$m and $\xi= 0.3~\mu$m in our numerical study. An initial vortex state is prepared by randomly choosing $N_v=32$ with $|s_i|=1$ and $\sum s_i =0$, where the distance to the nearest neighbor vortex is constrained to be larger than $7 \xi$ so as to prevent an unexpected, initial rapid decrease in the vortex number. In a regular vortex distribution, the intervortex distance is about $R/\sqrt{N_v}\sim 45\xi$. Figure 3 displays an example for the time evolution of a vortex state, where the vortex number decreases as the evolution proceeds. 

Figure 4(a) shows the decay curves of $N_v(t)$ obtained for various values of $\alpha$ [Fig.~4(a)]. We used 40 different initial states for statistical averaging. The vortex number shows a nonexponential decay behavior, which is also well described by the rate equation in Eq.~(1). The half decay time $t_h$ decreases with increasing $\alpha$, as expected, and the vortex decay rate $\Gamma=1/t_h$ is found to be linearly proportional to $\alpha$, as observed in the experiments [Fig.~4(b)]. The proportionality constant is measured to be $\Gamma/\alpha\approx 38$~s$^{-1}$. It is remarkable that the vortex decay rates obtained from the numerical simulations show good quantitative agreement with the experimental results. Recalling the limitations of our point-vortex model, this quantitative agreement should be taken with caution. Nevertheless, it appears that our point-vortex model reasonably captures the vortex dynamics of turbulent BECs.

To understand the linear relation between $\Gamma$ and $\alpha$, it is helpful to consider the motion of a single vortex dipole in a homogeneous system under thermal friction. A vortex dipole with a pair separation $d$ propagates linearly with the velocity of $v=\hbar/(m d)$. Because of thermal friction, the separation of the two vortices decreases as $\dot{d}(t) = -\alpha v =- \alpha \hbar /(m d)$ according to Eq.~(6), and eventually the vortex dipole will be annihilated after the time $\tau= m/(2\alpha \hbar) (d^2-d_c^2)$ for $d(\tau)\leq d_c$. If we regard a turbulent BEC as a gas of vortex dipoles with mean pair separation of $\bar{d} \gg d_c$, then the vortex decay rate may be estimated as $\Gamma\sim \tau^{-1}\approx 2 \alpha \hbar/(m \bar{d}^2)$. This is consistent with the observed linear dependence of $\Gamma$ on $\alpha$. Furthermore, for the case of a condensate with $N_v=32$ vortices in a cylindrical trap of radius $R=76~\mu$m, we have $\Gamma/\alpha \sim (\hbar/m) (N_v/R^2)\approx 15$~s$^{-1}$ with $\bar{d}\sim R/\sqrt{N_v}$, which is quite compatible with the value of $\Gamma/\alpha$ observed in the numerical simulations.

\begin{figure}
\includegraphics[width=8.5cm]{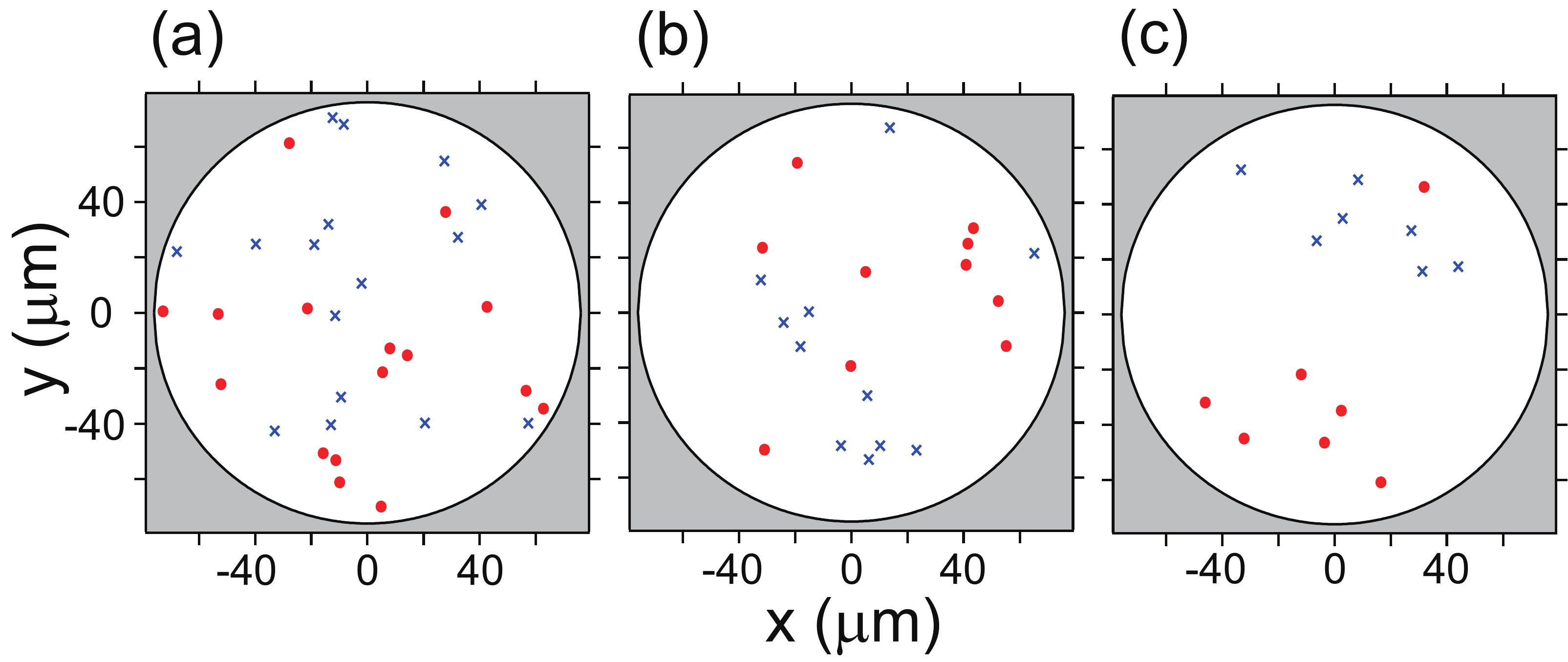}
\caption{(Color online) Emergence of same-sign vortex clusters without thermal friction. Time evolution of the vortex distribution at $t=$ (a) $0~$s, (b) $200~$s, and (c) $500~$s for $\alpha=0$. The initial vortex distribution is the same as that in Fig.~3(a).}
\end{figure}

\subsection{Effect of friction on vortex clustering}

It is known that in a point-vortex model, vortex states above a critical vortex interaction energy $E_c(N_v)$, referred to as negative-temperature states, would evolve into a large vortex structure where like-sign vortices are clustered~\cite{Yatsuyanagi05,Onsager49}. It was argued that vortex-antivortex pair annihilation would reinforce the vortex clustering behavior because  the energy of the vortex system does not change significantly after pair annihilation but the critical energy $E_c$ is lowered for smaller $N_v$~\cite{Simula14}. However, as demonstrated in previous numerical works using the damped GP equation~\cite{Billam15}, thermal dissipation would weaken the clustering tendency by decreasing the system energy below $E_c$. Emergence of vortex clustering might be critically determined by a competition between the two effects in the vortex dynamics.

In our simulations, we found that some of our 40 initial vortex states evolved into a state with two like-sign vortex clusters in the absence of thermal friction. Figure 5 displays an example for the time evolution with $\alpha=0$. Two clusters of same-sign vortices appear after a long evolution time with decreased vortex number [Fig.~5(c)]. Remarkably, we observed that the vortex clustering can be suppressed even by very small thermal friction. Indeed, Figure 3 shows the time evolution of the same initial vortex state for $\alpha=0.0025$, where the vortex positions are maintained in a random configuration over the evolution. This observation clearly demonstrates the adverse effect of thermal friction on the emergence of vortex clustering. In the experiment of Kwon~$\it{et~al.}$~\cite{Kwon14}, the vortex-clustered state could be suppressed in decaying turbulence because the value of $\alpha$ was estimated to be about 0.01 at the lowest temperatures, indicating that the turbulent BECs were in a strong dissipation regime, not allowing for the emergence of a vortex-clustered state out of decaying turbulence.

\section{Summary}

We investigated the role of thermal friction in the relaxation dynamics of turbulent BECs. By examining the correlation between the vortex decay rate $\Gamma$ and the thermal friction coefficient $\alpha$, which have been separately measured in recent experiments~\cite{Kwon14,Moon15}, we observed that $\Gamma$ is almost linearly proportional to $\alpha$. We performed numerical simulations of turbulent BECs using a dissipative point-vortex model and observed that the linear dependence of $\Gamma$ on $\alpha$ is quantitatively accounted for by the model. Furthermore, the simulation results showed that thermal dissipation in the experiment of Kwon~$\it{et~al.}$~\cite{Kwon14} was too strong to observe vortex clustering in decaying turbulence. Ensuing important questions are what is the temperature requirement for observing vortex clustering and whether it can be achievable in current experiments. In many numerical studies using the GP equation, thermal dissipation was taken into account by introducing a phenomenological damping parameter $\gamma$ and its value was estimated in a wide range of $10^{-1}$ to $10^{-4}$  for typical experimental conditions~\cite{Madarassy08,Rooney10, Bradley12,Neely13,Groszek16}. It is highly desirable to improve our quantitative understanding of the relation between $\gamma$ and  $\alpha$ as well as the temperature dependence of $\gamma$.

\begin{acknowledgments}
This work was supported by the Research Center Program of IBS (Institute for Basic Science) in Korea (IBS-R009-D1).
\end{acknowledgments}


\begin{references}
\bibitem{footnote1} We checked that the vortex decay rate is insensitive to the pair annihilation distance, $d_c$, up to $8\xi$.
\end{references}
\end{document}